\documentclass[amssymb,aps,twocolumn,nofootinbib]{revtex4}
\usepackage[]{graphicx,amsmath}

\def\ket#1{| #1 \rangle}
\def\bra#1{\langle #1 |}
\def\bracket#1#2{\langle #1 | #2 \rangle}
\def\II{1\!\mathrm{l}}
\newtheorem{theorem}{Theorem}
\newtheorem{lemma}{Lemma}

\begin{document}

\title{Classicality of quantum information processing}
\author{David Poulin\footnote{e-mail address: poulinda@iro.umontreal.ca}}
\affiliation{
D\'{e}partement de physique et DIRO, Universit\'{e} de Montr\'{e}al,
Montr\'{e}al, Canada H3C 3J7. }
\date{\today}

\begin{abstract}
The ultimate goal of the classicality program is to quantify the
amount of {\it quantumness} of certain processes. Here, classicality is
studied for a restricted type of process: quantum information processing (QIP).
Under special conditions, one can force some qubits of a
quantum computer into a classical state without affecting
the outcome of the computation. The minimal set of conditions is described and
its structure is studied. Some implications of this formalism are the
increase of noise robustness, a proof of the quantumness of mixed
state quantum computing and a step forward in understanding the very
foundation of QIP.
\end{abstract}

\maketitle

\section{Introduction}
\label{int}
It is known thanks to Bernstein and Vazirani \cite{BV1993} that,
in a realistic context where errors occur and finite probability of
success is accepted, quantum mechanics
offers greater computational power than classical mechanics for some oracle
types of problem. It was then shown by
Simon \cite{simon1994} that this gain can be exponential. While these
results are of great fundamental 
significance, it is the factoring algorithm by Shor
\cite{shor1994_1997} that has attracted 
the most interest even though it lacks a proof for the lower bound
of its classical
counterpart. Indeed, it is still an open question whether the laws of quantum
mechanics can increase our computational power in the absence of an
oracle.

On the other hand, physical implementation of these quantum
algorithms is extremely challenging due to the impossibility of
perfectly isolating a quantum system from the rest of the
universe \cite{zurek1991_landauer1995_unruh1995}. The accuracy threshold theorem
\cite{AB1997_kitaev1997_preskill1998,KLZ1998} indicates that quantum error correction codes
\cite{shor1995_KL1997_steane1996,gottesman1996} 
and fault tolerant quantum computing \cite{shor1996,KLZ1998,gottesman1996}
offer a theoretical solutions to
the decoherence problem, but the technical challenge remains. 

Hence, when using quantum mechanics for computation, one
would wish to make sure that no classical system can achieve the same
task short of necessitating an infeasible amount of
resources---either time, space, or energy. It therefore
seems natural to ask the question ``are we wasting 
quantum resources while computing?'' given a certain algorithm. A
negative answer would provide an answer for the previously mentioned
open question. A positive answer is more subtle since it is
possible to waste {\it part} of the quantum resources and judiciously
use the rest, hence achieving a speeding up over classical
computing. The part of a quantum algorithm that can be substituted by
classical information processing (CIP) without appreciably slowing down the
computation is what we here refer to as its {\it classicality}. If no such
substitution can be performed, we would be forced to conclude that the
algorithm is {\it fully  quantum}, once again answering the open
question.

Different approaches can be used to substitute quantum information
processing (QIP) by CIP\@. Feynman 
anticipated that the straightforward way---that is, to simply simulate
the quantum dynamics with a classical computer---would fail
\cite{feynman1982}. Indeed, 
this seems reasonable since the matrices we use to describe the evolution
and the state of the system grow exponentially. Nevertheless, this
intuition still has not been proved; keeping track of these huge
matrices might not be necessary. However, D. Aharonov and Ben-Or have 
shown that the classical simulation
of a quantum system can become efficient when decoherence is brought
into the picture \cite{AB1996}. This transition between a best known exponential and a
polynomial simulation cost depends on the decoherence rate $\eta$. Roughly
speaking, decoherence restricts the state of the system to a very small
part of the accessible states. The system collapses at a rate $\eta$ to
the (classical) pointer states \cite{zurek1981}; there is no large
scale coherence. Therefore, the classical simulation
becomes efficient since it has to deal with only polynomial-size
submatrices (representing the space ``close'' to the pointer states).
These facts suggest a measure of classicality based on
the ``volume'' of the total state space that is exploited by the system
during the computation. The smaller this volume, the easier it will be
to classically simulate the system. 

A second approach to the classicality problem is complexity
theory \cite{BV1997}. One can try to reduce the complexity of a
quantum algorithm 
by using a hybrid classical-quantum computer. Shor's algorithm is an
example of such a hybrid computer.  Cleve and Watrous have further
exploited this idea and reduced the depth of the factoring algorithm
from polynomial to logarithmic \cite{CW2000}. This depth is achieved at the price of
polynomial pre- and post-CIP. A measure of the {\it quantumness}---as
opposed to classicality---would then be the complexity of the 
quantum part of the algorithm optimized for a hybrid computer. 

The approach I present here is radically different. Parts of the
system---some qubits 
of the quantum computer---are forced to classical states. This
``forcing'' is done in a way that will not affect the result of the 
computation. Hence, the quantum information conveyed by
these qubits can be replaced by classical information,
until a new transformation makes them quantum again. Thus, the
computer becomes a classical-quantum hybrid. Nevertheless, it is on a
smaller scale than the Cleve-Watrous example since this procedure is
applied at every step of the computation. In contrast with
the approach of Aharonov and Ben-Or---where the environment would
decrease the performance of QIP so it could be simulated by CIP---my
approach does not affect the computation even if it forces parts of
the computer to become classical. 

While the above was presented informally, the next
section introduces the formalism used to specify how and when one can
force the state of some qubits to become classical. Unfortunately, the
conditions yielded by this formalism are very cumbersome, so Sec. \ref{analysis}
deals with lemmas and theorems that might help simplify the
analysis.  I present in Sec. \ref{results} an academic example where the
formalism is applied. I also relate the concept of classicality to
noise robustness and computing with mixed states. I conclude in
Sec. \ref{conclusion} with some consequences of this work.

\section{Formalism}
\label{formalism}

The formalism used to describe classicality is the one of consistent
histories (CH's), first introduced in the foundation of quantum mechanics
by Griffiths 
\cite{griffiths1984}, later investigated and modify by Omn\`{e}s
\cite{omnes1988} and Gell-Mann and Hartle \cite{GH1990}. While
this theory is still much debated, its use in QIP does not suffer
from any of the usual criticism. The theory was originally intended as
an interpretation of quantum mechanics where the relation between the
classical and the quantum world would be clear. The main obstacle to
such a formalism is the lack of a precise description of the classical
world. It is indeed quite difficult to write down the equations that
characterize something we cannot describe.  Since CIP has a precise
description, this obstacle will be avoided. 

\subsection{General formalism}

The goal of the formalism is to select the quantum histories---time
ordered sequences of projection operators---to which a 
{\it good} probability distribution can be assigned.
We call $\hat\sigma =\{\hat P_\alpha\}_{\alpha=1,\ldots ,m}$ an exhaustive set of 
exclusive projectors if it satisfies
\begin{equation}
\sum_{\alpha =1}^m \hat{P}_{\alpha}=\II \ \ \  \text{and} \ \ \ 
\hat{P}_{\alpha}\hat{P}_{\beta}=\delta_{\alpha\beta}\hat{P}_\alpha.
\label{proj}
\end{equation}
Such  decompositions of the identity operator are independently made at $n$ different
times $t_1 < t_2 < \ldots  < t_n$, hence defining $n$ sets of projectors
$\sigma^{(j)}(t_j) = \{P^{(j)}_{\alpha_j}(t_j)\}_{\alpha_j =1,\ldots ,m_j}$ that 
are now written in the Heisenberg picture, $P^{(j)}_{\alpha_j}(t_j) =
   U^\dag(t_j,0)\hat{P}^{(j)}_{\alpha_j}U(t_j,0)$, where $U$ is the
evolution operator. The explicit time label
will henceforth be dropped since this causes no possible
confusion. Notice that the cardinality of each set is independent, so
the rank of every projector is arbitrary as long as eq.(\ref{proj}) is
satisfied. The basic ingredient of the CH formalism is, of course, a
quantum history. A history is 
constructed by picking one projector from each of the $n$ sets. With
every history is associated a history operator $C_\alpha =
   P^{(1)}_{\alpha_1} P^{(2)}_{\alpha_2}\ldots  P^{(n)}_{\alpha_n}$ where the
label $\alpha$ stands for $(\alpha_1, \alpha_2, \ldots ,\alpha_n )$. An
exhaustive family of disjoint histories is obtained by choosing one
projector in each of the $n$ sets in every possible way. We will
denote such a family ${\mathcal S}=\{\rho, \sigma^{(1)}, \sigma^{(2)},\ldots ,
   \sigma^{(n)}\}$ where $\rho$ is the initial state of the
system. ${\mathcal S}$ contains 
\begin{equation}
N=\prod_{j=1}^n m_j
\label{N}
\end{equation}
histories. It is
said to be exhaustive because the $N$ history operators sum to identity. The
histories making up ${\mathcal S}$ are said to be disjoint because they all
differ at least at one time by construction. 

The probability of a history $\alpha$---the probability that the
physical system starting in state $\rho$ at time $0$ is in the
spectrum of $P^{(1)}_{\alpha_1}$ at time $t_1$ and in the spectrum of 
$P^{(2)}_{\alpha_2}$ at time $t_2$ and so on.---is given by
\begin{eqnarray}
Pr(\alpha)&=& Pr(\alpha_1, \alpha_2, \ldots , \alpha_n) \nonumber\\
&=&Tr\left\{ P^{(n)}_{\alpha_n}\ldots P^{(2)}_{\alpha_2} P^{(1)}_{\alpha_1} \rho
P^{(1)}_{\alpha_1} P^{(2)}_{\alpha_2}\ldots   P^{(n)}_{\alpha_n} \right\} \nonumber\\
&=&Tr\left\{C_{\alpha}^{\dag}\rho C_{\alpha}\right\},
\end{eqnarray}
which can easily be derived from the standard Copenhagen
interpretation. Another fact that has been learned from elementary
quantum mechanics is 
that it is usually forbidden to assign probabilities to quantum histories
(recall Young's slit experiment). The CH formalism yields conditions
under which the selected histories---the {\it consistent
  histories}---can be assigned probabilities without any risk of logical
contradiction. 

The coherence function that maps $history \times history \mapsto
   \mathbb{C}$ is defined by
\begin{equation}
D(\alpha;\beta)=
Tr\left\{C_{\alpha}^{\dag}\rho C_{\beta}\right\}.
\label{coherence}
\end{equation}
Roughly speaking, $D(\alpha ;\beta)$ is the average interference
between the Feynman paths following history $\alpha$ and those of
history $\beta$ \cite{GH1990}.
The necessary and sufficient condition for which a family of histories
can be used to describe the evolution of a quantum system without
logical contradictions is \cite{griffiths1984}
\begin{equation}
Re\left[D(\alpha;\beta)\right]=\delta_{\alpha\beta}Pr(\alpha)
\ \ \ \forall\ \alpha,\beta \in {\mathcal S}
\label{weak}
\end{equation}
where $Re$ is the real part and $\delta_{\alpha\beta}$ stands for 
$\delta_{\alpha_1\beta_1}
   \delta_{\alpha_2\beta_2}\ldots  \delta_{\alpha_n\beta_n}$. Eq.(\ref{weak})
is known as the {\it weak consistency condition}.\footnote{Some
  authors would claim that this condition should only be imposed to
  histories $\alpha$ and $\beta$ such that $C_\alpha +C_\beta$ is also
  an chain of projectors. I do not wish to enter this debate here, see
  additional note in \cite{GH1994} for more details. Furthermore, all
  possibles ambiguities will disappear with the introduction of the
  computing consistency condition (eq.\ref{comp}).} One can think of
this condition as an insensibility of the system to the measurements
described by the $\sigma$'s making up the consistent family. Whether
the measurement $\sigma^{(k)}$ is carried out at time $t_k$ or not will not
influence the statistical outcome of the measurement $\sigma^{({k'})}$ at
time $t_{k'}$. This does not mean that the measurements leave the
state of the system unchanged. In general, the system will collapse to a
different state but in a way that does not affect the statistical
results. 

Obviously, depending on the context, the strict imposition of this
condition might not be necessary. One can be satisfied if the real
parts of the off-diagonal terms of the coherence function are sufficiently
small, something known as the $\epsilon$-consistency. In this case,
classical logic would be valid up to some finite accuracy.

Since logic can be applied to the classical world, weak consistency is
a necessary condition for classicality. Nevertheless, one can find
many examples that are obviously in a quantum regime while satisfying
condition eq.(\ref{weak}). For this reason and to facilitate some analysis,
more restrictive conditions
have been introduced, such as the {\it medium consistency condition}
\cite{GH1990}:
\begin{equation}
D(\alpha;\beta)=\delta_{\alpha\beta}Pr(\alpha)
\ \ \ \forall\ \alpha,\beta \in {\mathcal S}
\label{medium}
\end{equation}
where we have simply dropped the $Re$. Once again, it was soon
realized 
that this condition is not sufficient to separate the classical
regime. To my knowledge, no counterexample has been found to the
{\it strong consistency condition} \cite{GH1995} which is defined as 
\begin{equation}
\forall \alpha \in {\mathcal S},\ \exists R_\alpha \ : \ \ 
C_\alpha \rho = R_\alpha \rho
\end{equation}
where the $R_\alpha$ are projection operators. This last definition is
intimately related to decoherence since it implies the existence of a
record of the system's history \cite{GH1993}. 

Suppose that ${\mathcal S}=\{\rho,\sigma^{(1)}, \ldots , \sigma^{(n)}\}$ forms a family of
CH's \footnote{Most of the following definitions are from
  \cite{DK1996}.}. The family ${\mathcal S'}=\{\rho,\sigma^{(1)},
   \ldots ,\sigma^{(k)},\tau,\sigma^{({k+1})}\ldots , 
   \sigma^{(n)}\}$ will be called a consistent extension of ${\mathcal S}$ by the set of
projectors  $\tau=\{Q_i\}_{i=1,\ldots ,m}$ if $\tau$ satisfy
condition (\ref{proj}) and ${\mathcal S'}$ is consistent.
Here, we will consider only the case $k<n$; the final measurements are
fixed. A consistent
extension will be said to be trivial when for every history
$(P^{(1)}_{\alpha_1}, \ldots ,  P^{(n)}_{\alpha_n})$ in ${\mathcal S}$ there is at most one
history $(P^{(1)}_{\alpha_1}, \ldots ,
   P^{(k)}_{\alpha_k},Q_i,P^{(k+1)}_{\alpha_{k+1}}, \ldots , P^{(n)}_{\alpha_n})$ in
${\mathcal S'}$ that has a nonzero probability. One could say that the
measurement $\tau$ does not yield any new information since $Pr(Q_i |
   \alpha_1, \alpha_2, \ldots , \alpha_n)$ is a deterministic process. 

When all projectors in $\sigma'$ are sums of projectors in $\sigma$,
we say that $\sigma'$ is a coarse graining of $\sigma$. One can
easily verify that coarse graining preserves consistency, a
consequence of the classical probability sum rules.
The  set $\sigma$ will be said fine grained if it is not
the coarse graining of  any set, {\it i.e.} if all its projectors have
rank 1. 

\subsection{Applying CH formalism to QIP}

A quantum algorithm is the specification of an initial state $\rho$,
an evolution operator $U(t)$, and a final measurement
$\sigma^{(f)}$. Hence, it can be seen as a one-event ($n=1$) family of consistent
histories ${\mathcal S}=\{\rho,\sigma^{(f)}\}$ \footnote{A one-event family is
always consistent.}. Since CH's are good candidates to
describe the classical world, our goal will be to make consistent
extensions of this one-event family, hence describing the quantum
algorithm in classical terms as much as possible. Since a quantum
algorithm is described using discrete unitary evolutions (gates), the
set of discrete times $t_i$ has a very natural definition. 

The difficulty of defining a consistency condition that is {\it sufficient}
for classicality will be avoided in QIP by adding an extra condition :
\begin{center}
{\it The consistent extensions must be made in a local basis}.
\end{center}
Furthermore, the condition eq.(\ref{weak}) is indeed quite weak, but in
the context of QIP it is natural to consider a weaker one. This new
condition, although it has not yet found a concrete application, is
the weakest one that allows for an efficient classical
simulation. This condition is based on the fact that the outcome of
the intermediate measurements---the ones used to force parts of the
quantum computer to classical states---is of no interest. Only the
statistical outcome of the final measurement is relevant. I call the
consistency condition obtained from such a requirement the {\it
computing-consistency condition}:
\begin{equation}
\sum_{\substack{\alpha \\ \alpha_n = k}} 
\sum_{\substack{\beta < \alpha \\ \beta_n = k}}
Re\left[D(\alpha;\beta)\right] =0
\ \ \ \forall \ k=1,\ldots ,m_n.
\label{comp}
\end{equation}
This reduces the number of conditions from $N(N+1)/2$ [eq.(\ref{N})] to $m_n$.
It is straightforward to verify that the consistency conditions form a
hierarchy 
\begin{equation}
Strong\ \Rightarrow\ Medium\ \Rightarrow\ Weak\ \Rightarrow\
Computing.
\label{hierarchy}
\end{equation}

The restriction to local sets of projectors together with
(eq. \ref{comp}) yields a
necessary and sufficient set of conditions for CIP. Indeed, if
complete local sets of projectors $\sigma^{(k)}$ can be inserted
consistently between each gate $U_k$ and $U_{k+1}$ of a quantum
circuit, the {\it effective} dynamics of the quantum system can be
simulated by classical spins undergoing stochastic evolution. Note
that the dynamics of the quantum system is not the same as that of
the classical spins; it is only the dynamics {\it as seen from the
measurements} making up the family of histories that is the same,
and hence a simulation of the ``effective'' dynamics.  

To illustrate this, assume that a quantum algorithm is of the form 
\begin{equation}
{\mathcal Q} = U_n \ldots U_2 U_1
\end{equation}
with initial state $\ket{\psi_0}$ and final measurement $\ket{\phi_i^{(n)}}$,
where $U_i$ is the unitary operator applied at step $i$. If $\mathcal
Q$ admits complete local computational consistent extensions 
$\hat\sigma^{(k)} = \{\ket{\phi^{(k)}_{j}}\bra{\phi^{(k)}_j}\}$, we
get the following equality: 
\begin{eqnarray}
&&|\bra{\phi^{(n)}_i} U_n \ldots U_2 U_1 \ket{\psi_0}|^2 \nonumber\\
&=& \sum_{\alpha}\bra{\psi_0}U_1^\dagger P_{\alpha_1}^{(1)}\ldots
P_{\alpha_{n-1}}^{(n-1)} U_n^\dagger \ket{\phi_i^{(n)}}\nonumber\\
&\times&\sum_\beta\bra{\phi_i^{(n)}} U_n
    P_{\beta_{n-1}}^{(n-1)} \ldots P_{\beta_1}^{(1)} U_1\ket{\psi_0}
\nonumber \\
&=&(\phi^{(n)}_i|{\mathcal T}_n \ldots{\mathcal T}_2 {\mathcal T}_1
|\phi^{(0)}_j) \ |(\phi^{(0)}_j\ket{\psi_0}|^2 ,
\label{stoch}
\end{eqnarray}
where $(\mathcal T_k)_{ij} =
|(\phi^{(k)}_i|U_k|\phi^{(k-1)}_j)|^2$ are stochastic
transition matrices and
the $|\cdot )$ notation is to emphasize that these are classical
states.\footnote{The classical state of a qubit can be represented by
  a vector pointing in a Bloch sphere. The $\epsilon$-consistency
  handles the finite accuracy issue.} The first equality is a consequence of the exhaustiveness of the
family of histories while the second equality simply follows from
eq.(\ref{comp}). 
Therefore, the final result can be obtained by replacing each quantum
unitary evolution on a quantum superposition by a stochastic evolution
on a classical mixture.

Another resource that can be exploited to facilitate the substitution
of QIP by CIP is
feed-back.  We can formalize this idea with {\it branch-dependent CH's}
\cite{omnes1988}. To force some qubits into classical states, we have measured
them in an appropriate basis. The acquired information can allow us 
to make the quantum algorithm even more classical. Indeed, Paz and
\.Zurek \cite{PZ1993} have illustrated a physical system whose evolution
could be described classically in a consistent fashion only in the
presence of feedback. In this setting, 
the set of projector $\sigma^{(j)}$ at time $t_j$ is chosen according
to the outcome of the previous measurements $\alpha_1, \alpha_2, \ldots 
   ,\alpha_{j-1}$. We thus write
\begin{equation}
\sigma^{(j,\alpha_1,\alpha_2,\ldots \alpha_{j-1})}
=\left\{P_{\alpha_j}^{(j,\alpha_1,\alpha_2, \ldots , \alpha_{j-1})}
\right\}.
\end{equation}
It is interesting to note for future analysis that branch-dependent CH's
are equivalent to ``normal'' CH's with ancillary qubits \footnote{ I thank
Charles Bennett for pointing this out to me.}. In the context of QIP,
the processing required for the feedback should be restricted to
polynomial time and space for obvious reasons.

\section{Classicality analysis}
\label{analysis}

We now have a very clear statement of the classicality problem: Given
a quantum algorithm, we must find sets of local projectors that, when
applied between the gates, constitute a consistent extension of the original
one-event family. This task is in no sense trivial. In
this section, I present new results on the structure of CH's. The goal
is twofold. On one hand, they might help in finding bounds on the
classicality of a quantum algorithm, that is, the maximum amount of QIP
that can be substituted by CIP. I do not claim that this goal is
attained, but the following results give good starting points. 
On the other hand, these results suggest a 
simplified way of analyzing an algorithm from a CH point of view. This
section should be seen as a complement to a similar study by Dowker 
and Kent \cite{DK1996}. 

The first result concerns the number of nontrivial consistent
extensions that can be made for a general system using the medium
consistency condition eq.(\ref{medium}). Part one of this result was 
first proven by Di\'{o}si \cite{diosi1994}.

\begin{lemma}
Given an initial density matrix $\rho$ of rank $r$, there can be no
more than $r\Omega$ medium CH's with nonzero probability in a
family, where $\Omega$ is the dimension of the Hilbert space.
On the other hand, there are infinitely many families that
attain this bound. 
\label{diosibound} 
\end{lemma}

\noindent {\it Proof}. Part one of the lemma is Di\'{o}si's bound. To
construct a family that attains this bound, choose $\sigma^{(1)}$ as the 
set of rank 1 projectors that commute with $\rho$---that is, $\rho =
   \sum p_i \ket{\psi_i}\bra{\psi_i}$ and
$\sigma^{(1)} =\{\ket{\psi_i}\bra{\psi_i}\}$---and
$\sigma^{(2)}=\{\ket{\phi_i}\bra{\phi_i}\}$ with 
$\bracket{\phi_i}{\psi_j} \neq 0 \ \forall\ i,j$. For example, this can
be done by letting $\ket{\phi_j}$ be the Fourier transform 
\begin{equation}
\label{fourier}
\mathcal F \ket{x} =
\frac{1}{\sqrt{\Omega}}\sum_{y=0}^{\Omega-1}e^{i2\pi xy/\Omega}\ket{y}
\end{equation}
of $\ket{\psi_j}$. Then, $S=\{\rho,\sigma^{(1)},\sigma^{(2)}\}$ is
consistent and 
has exactly $r\Omega$ nonzero probability histories. From this
example, it is 
obvious that infinitely many families can be constructed.
\hfill$\square$   

\medskip

The reason why Di\'{o}si's bound was reached in Lemma \ref{diosibound}
is that $\bracket{\phi_i}{\psi_j} \neq 0 \ \forall\ i,j$. Without this
condition, it is clear that Di\'{o}si's bound is not attained. The next
result shows that in that case no
consistent extension will ever reach Di\'{o}si's bound either.

\begin{lemma}
Let ${\mathcal S}=\{\rho ,\sigma^{({1})},\sigma^{({2})}\}$ where $\sigma^{(1)}$ is chosen to be
the set of rank-1 projectors diagonal in the eigenbasis of $\rho$ and 
$\sigma^{(2)}$ is any set of rank-1 projectors. Then, ${\mathcal S}$ is medium consistent and 
there are no sets of projectors
$\tau^{(1)}, \ldots , \tau^{(n)}$ that make ${\mathcal S'}=\{\rho ,\sigma^{({1})},\tau^{(1)}, \ldots , \tau^{(n)}
   ,\sigma^{({2})}\}$ a nontrivial consistent extension of ${\mathcal S}$.
\label{interpolation}
\end{lemma}

\noindent {\it Proof}. First, we must show that ${\mathcal S}$ is consistent. This can
be done directly by writing
$\sigma^{(1)}=\{\ket{\psi_{i}}\bra{\psi_{i}}\}$ with 
$\rho=\sum_i \lambda_i \ket{\psi_i}\bra{\psi_i}$ and
$\sigma^{(2)}=\{\ket{\phi_{j}}\bra{\phi_{j}}\}$. We get
\begin{equation}
D(i,j;k,l) 
=\lambda_{i}\left|\bracket{\phi_{j}}{\psi_{i}}\right|^2
\delta_{ik} \delta_{jl},
\end{equation}
which is exactly the medium consistency condition. Second, suppose 
that $\tau^{(j)}=\{P^{(j)}_{\alpha_j}\}$ are sets of projectors that make
${\mathcal S'}$ consistent.
For a given couple $(\ket{\psi_{i}}, \ket{\phi_{j}})$
the consistency condition eq.(\ref{medium}) reads
\begin{eqnarray*}
&&D(i,\alpha_1, \ldots , \alpha_n, j; i,\beta_1, \ldots ,
  \beta_n, j) \\
&=& \bra{\phi_j}P^{(n)}_{\alpha_n}\ldots
P^{(1)}_{\alpha_1}\ket{\psi_i}\bra{\psi_i}  
P^{(1)}_{\beta_1}\ldots P^{(n)}_{\beta_n}\ket{\phi_j}\lambda_i \\
&=& \delta_{\alpha_1 \beta_1}\ldots \delta_{\alpha_n \beta_n}Pr(i,\alpha_1, \ldots ,
\alpha_n, j) \\
&\varpropto& \sqrt{Pr(i,\alpha_1,\ldots , \alpha_n, j)}\times
 \sqrt{Pr(i,\beta_1,\ldots , \beta_n, j)}
\end{eqnarray*}
so for this given couple there is just one extension that has nonzero
probability, proving the lemma.\hfill$\square$

\medskip

The next result is the adaptation of Di\'{o}si's bound to the weak
consistency condition. One must keep in mind that when CH's are applied
to QIP one should always use the weakest condition possible.
Interestingly, this bound imposes a limitation on the amount
of information that can be extracted on the evolution of a quantum
system and on the number of Everett branches \cite{everett1957} in the
universe. Indeed, since every branch has to obey its own logic, one
can argue that the branching must be made in a consistent fashion
\cite{DK1996}. Paz and \.Zurek's ``{\it decoherence defines branches}''
\cite{PZ1993} can be reformulated as {\it consistency defines
  branches} since decoherence implies consistency.

\begin{lemma}
Given an initial density matrix $\rho$ of rank $r$, there can be no
more than $2r\Omega$ weak CH's with nonzero probability in a family. On
the other hand, there are infinitely many families that attain this
bound. 
\label{diosibound2}
\end{lemma}

\noindent {\it Proof}. Part one of the proof is very similar to
Di\'{o}si's proof. Let $\{ p_i,\ket{\psi_i}\}$ be the eigenvalues and
eigenstates of $\rho$. Define the purification of $\rho$ 
\begin{equation}
\ket{\Psi}=\sum_{i=1}^{r}\sqrt{p_i}\ket{\psi_i}\otimes\ket{\psi'_i}
\end{equation}
where the states $\ket{\psi'_i}$ form a basis of a Hilbert
space $\mathcal{E}$ of dimension $r$, so
$\rho=Tr_{\mathcal{E}}\{\ket{\Psi}\bra{\Psi}\}$. Now, define the
non-normalized states $\ket\Psi_{\alpha}=(C_{\alpha}\otimes\II)
   \ket{\Psi}$. The weak consistency condition becomes
$Re\bracket{\Psi_{\alpha}}{\Psi_{\beta}}=\delta_{\alpha
   \beta}Pr({\alpha})$. Since these states lie in a Hilbert space of
dimension $\Omega r$, there can be no more than $2\Omega r$ nonzero
such vectors. 
For the second part of the proof, we will assume that $\Omega$ is
even, the odd case being more technical and nonpertinent for
QIP. Choose $\sigma^{(1)}$ as above  
and $\sigma^{(2)}$ as the Fourier transform of $\sigma^{(1)}$;
$\sigma^{(3)}$ is 
obtained by applying the transformation
\begin{equation}
\ket{j} \mapsto \frac{i\ \ket{j}+\ket{(j+\Omega/2){\mathsf{mod}}
    \Omega}}{\sqrt{2}}
\label{transfo}
\end{equation}
to every element of $\sigma^{(2)}$ (for systems composed of qubits,
this is simply the gate 
\begin{equation}
\frac{1}{\sqrt{2}}\left(
\begin{array}{cc}
i & 1 \\
1 & i
\end{array}\right)
\end{equation}
 on the most significant qubit). The
consistency of this family will easily be verified after the main
result of this section is established. The
rest of the proof is identical to lemma \ref{diosibound}.
\hfill$\square$

\medskip

I have not found an adaptation of lemma \ref{interpolation} to the
weak consistency condition. One should always try to adapt the results to
the weakest condition---the computing-consistency condition---which
would make them general. Unfortunately, this is not always simple.
 
One of the main difficulties of the classicality problem stated at
the beginning of this section is the infinite number of sets of
projectors that are candidates for consistent measurements. The
following result considerably restricts these candidates. Indeed, it shows
that searching among the sets of {\it rank}-1 projectors is general,
at least for the medium consistency condition.

\begin{theorem}
Let ${\mathcal S}=\{\rho ,\sigma^{(1)}, \sigma^{(2)}, \ldots , \sigma^{(n)}\}$ be a family of
medium CH's with $\rho=\ket{\psi}\bra{\psi}$ a pure state. There exists a
consistent family  ${\mathcal S'}=\{\rho ,\sigma^{\prime (1)},
\sigma^{\prime (2)}, \ldots , \sigma^{\prime (n)}\}$ 
with
$\sigma^{\prime (i)}=\{\ket{\phi^{(i)}_j}\bra{\phi^{(i)}_j}\}$
being a fine graining of $\sigma^{(i)}$. 
\label{fine}
\end{theorem}

\noindent {\it Proof}. The proof is by induction. We first show that,
given  ${\mathcal S}$, we can replace the last set of projectors $\sigma^{(n)}$ by
one of its fine graining $\sigma^{\prime (n)}$ in a consistent manner. Then we
show that  if $\sigma^{(i)}$ are all fine grained for $i>k$ we can replace
$\sigma^{(k)}$ by  one of its fine graining $\sigma^{\prime (k)}$. 

\noindent Define $\ket{\psi_{\alpha_1, \ldots , \alpha_{n}}} \equiv
   P^{(n)}_{\alpha_{n}} \ldots P^{(1)}_{\alpha_1}\ket{\psi}$. The consistency
condition \ref{medium}  asserts that $\bracket{\psi_{\alpha_1,
   \ldots ,\alpha_{n}}}{ \psi_{\beta_1,
   \ldots ,\beta_{n}}}=\delta_{\alpha_1\beta_1}
   \ldots \delta_{\alpha_{n}\beta_{n}}Pr(\alpha_1, \ldots , \alpha_{n})$ so these
vectors are orthogonal and have norm $\sqrt{Pr(\alpha_1, \ldots ,
   \alpha_{n})}$. We change the multi-index notation $(\alpha_1, \ldots ,
   \alpha_{n-1})$ to a single index $i$ that
are given by decreasing order of probability $Pr(i,
   \alpha_n)$. Now, condition (\ref{proj}) tells us that
$P^{(n)}_{\beta_n}\ket{\psi_{i,\alpha_n}}= \delta_{\alpha_n
   \beta_n}\ket{\psi_{i,\alpha_n}}$: the $\ket{\psi_{i,\alpha_n}}$ are
orthogonal vectors contain in the $\alpha_n\!-$subspace spanned by
the projector $P^{(n)}_{\alpha_n}$. The dimension of this subspace
being $rank\{P^{(n)}_{\alpha_n}\}=r^{(n)}_{\alpha_n}$, we get the
inequality $cardinality\{i:\ket{\psi_{i,\alpha_n}}\neq
   0\}=\omega^{(n)}_{\alpha_n} \leq r^{(n)}_{\alpha_n}$, so for each
$\alpha_n$ we define the normalized kets $\ket{\nu_{i,\alpha_n}}$ 
\begin{equation}
\ket{\nu _{i,\alpha_n}} =\left\{
\begin{array}{cc}
\frac{\ket{\psi_{i,\alpha_n}}}{\sqrt{Pr(i,\alpha_n)}} & for \ i\leq\omega_{\alpha_n}^{(n)} \\
\ket{\mu_{i-\omega^{(n)}_{\alpha_n}}} & for \ \omega^{(n)}_{\alpha_n}<i<r^{(n)}_{\alpha_n}
\end{array}
\right.
\end{equation}   
with $\{\ket{\mu_{i}}\}_{i=1, \ldots ,r^{(n)}_{\alpha_n}-\omega^{(n)}_{\alpha_n}}$
being any set of kets that completes the $\alpha_n\!-$subspace. 
Using the fact that $\bracket{\nu_{i,\beta_n}}{\psi_{\alpha_1, \ldots ,
    \alpha_{n}}}=0 \ \forall i\geq\omega^{\beta_n}$, 
one can easily verify that
\begin{equation}
P^{(n)}_{\alpha_n}=\sum_{i=1}^{r^{(n)}_{\alpha_n}} \ket{\nu
  _{i,\alpha_n}} \bra{\nu _{i,\alpha_n}}
\end{equation}
and that ${\mathcal S'}=\{\rho,\sigma^{(1)}, \ldots , \sigma^{({n-1})},
\sigma^{\prime ({n})}\}$ with 
$\sigma^{\prime (n)}=\{\ket{\nu _{i,\alpha_n}}\bra{\nu _{i,\alpha_n}}\}$ is a
consistent family. 

\noindent  To complete the proof, assume that the consistent family is
of the form
  ${\mathcal S}=\{\rho, \sigma^{(1)}, \ldots , \sigma^{(n)}, \tau^{(n+1)}, \ldots ,
   \tau^{(f)}\}$ with $\tau^{(j)}=\{\ket{\phi^{(j)}_i}\bra{\phi^{(j)}_i}\}$. The
consistency condition asserts that
$\bra{\psi}P^{(1)}_{\alpha_1}\ldots P^{(n)}_{\alpha_{n}} 
   P^{(n)}_{\beta_{n}}\ldots 
   P^{(1)}_{\beta_1}\ket{\psi}=\delta_{\alpha_1\beta_1}
   \ldots \delta_{\alpha_{n}\beta_{n}}Pr(\alpha_1, \ldots , \alpha_{n})$
so we can define 
$\ket{\psi_{\alpha_1, \ldots , \alpha_{n}}}$,
and $\ket{\nu _{i,\alpha_n}}$ in the same way we did previously. For the
same reason, one can see that 
\begin{equation}
P^{(n)}_{\alpha_n}=\sum_{i=1}^{r^{(n)}_{\alpha_n}} \ket{\nu
  _{i,\alpha_n}}\bra{\nu _{i,\alpha_n}} 
\end{equation}
and that ${\mathcal S'}=\{\rho,\sigma^{(1)}, \ldots , \sigma^{({n-1})},
\sigma^{\prime (n)}, \tau^{(n+1)}, \ldots , \tau^{(f)}\}$ with 
$\sigma^{\prime (n)}=\{\ket{\nu _{i,\alpha_n}}\bra{\nu _{i,\alpha_n}}\}$ is a
consistent family, completing the proof. \hfill $\square$

\medskip 
A different way of stating this result is to say that every family of 
medium CH's starting in a pure state can be obtained from coarse graining a fine grained family.  
For fined grained histories, Griffiths has introduced the notion of
consistent trajectories \cite{griffiths1993}, in analogy with a
trajectory in the classical 
phase space. I present this concept here because it will be very
useful to analyze QIP. The basic idea is to replace the condition eq.(\ref{medium})
with a graph analysis. The graph is constructed in the following
way. Since we are only considering rank-1 projectors, every $\hat\sigma^{(j)}$
now represents a choice of basis $\{\ket{\phi^{(j)}_i}\}$, written in
the Schr\"{o}dinger picture. With every
basis vector is associated a vertex in the graph. Two vertices are
connected by an edge if and only if they are separated by one time interval and
\begin{equation}
G^{ij}(t_{k+1},t_k)=
\bra{\phi^{(k+1)}_i}U(t_{k+1},t_k)\ket{\phi^{(k)}_j}\neq 0;
\end{equation} 
this last quantity is called the Green function. 

To illustrate this
concept, we construct the graph associated with the family used in the
last part of the proof of Lemma \ref{diosibound2} (with $\Omega = 6$). 
\begin{figure}[!h]
\begin{center}
\begin{picture}(180,145)
\setlength{\unitlength}{0.8pt}
\put(30,20) {\circle*{5}}
\put(30,50) {\circle*{5}}
\put(30,80) {\circle*{5}}
\put(30,110) {\circle*{5}}
\put(30,140) {\circle*{5}}
\put(30,170) {\circle*{5}}
\put(90,20) {\circle*{5}}
\put(90,50) {\circle*{5}}
\put(90,80) {\circle*{5}}
\put(90,110) {\circle*{5}}
\put(90,140) {\circle*{5}}
\put(90,170) {\circle*{5}}
\put(150,20) {\circle*{5}}
\put(150,50) {\circle*{5}}
\put(150,80) {\circle*{5}}
\put(150,110) {\circle*{5}}
\put(150,140) {\circle*{5}}
\put(150,170) {\circle*{5}}
\put(210,20) {\circle*{5}}
\put(210,50) {\circle*{5}}
\put(210,80) {\circle*{5}}
\put(210,110) {\circle*{5}}
\put(210,140) {\circle*{5}}
\put(210,170) {\circle*{5}}
\put(30,80) {\line(1,0){60}}
\put(90,79.4) {\line(1,0){60}}
\put(90,80.6) {\line(1,0){60}}
\put(90,79) {\line(1,-1){60}}
\put(90,81) {\line(1,-1){60}}
\put(90,79.2) {\line(2,-1){60}}
\put(90,80.8) {\line(2,-1){60}}
\put(90,79.2) {\line(2,1){60}}
\put(90,80.8) {\line(2,1){60}}
\put(90,79) {\line(1,1){60}}
\put(90,81) {\line(1,1){60}}
\put(90,78.8) {\line(2,3){60}}
\put(90,81.2) {\line(2,3){60}}
\put(150,20) {\line(2,3){60}}
\put(150,50) {\line(2,3){60}}
\put(150,80) {\line(2,3){60}}
\put(150,110) {\line(2,-3){60}}
\put(150,140) {\line(2,-3){60}}
\put(150,170) {\line(2,-3){60}}
\put(15,20) {\makebox(0,0)[r]{$5$}}
\put(15,50) {\makebox(0,0)[r]{$4$}}
\put(15,80) {\makebox(0,0)[r]{$j=3$}}
\put(15,110) {\makebox(0,0)[r]{$2$}}
\put(15,140) {\makebox(0,0)[r]{$1$}}
\put(15,170) {\makebox(0,0)[r]{$0$}}
\put(30,180) {\makebox(0,0)[c]{\large$\hat\sigma^{(0)}$}}
\put(90,180) {\makebox(0,0)[c]{\large$\hat\sigma^{(1)}$}}
\put(150,180) {\makebox(0,0)[c]{\large$\hat\sigma^{(2)}$}}
\put(210,180) {\makebox(0,0)[c]{\large$\hat\sigma^{(3)}$}}
\linethickness{2pt} 
\put(150,20) {\line(1,0){60}}
\put(150,50) {\line(1,0){60}}
\put(150,80) {\line(1,0){60}}
\put(150,110) {\line(1,0){60}}
\put(150,140) {\line(1,0){60}}
\put(150,170) {\line(1,0){60}}
\end{picture}
\caption{Graph analysis of Lemma \ref{diosibound2}.}
\label{graphe_lemme}
\end{center}
\end{figure}
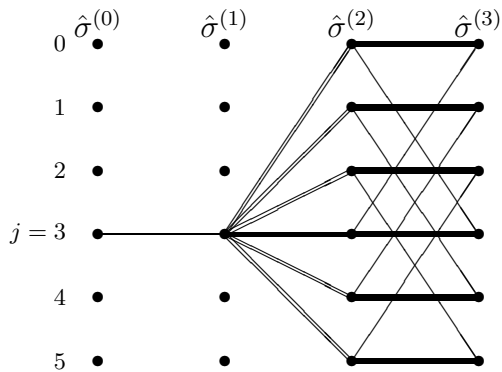
$\hat\sigma^{(0)}$ is the set of initial states $\ket{\psi_i}$---if
each of these states generates a consistent family, their statistical
mixture will also, by linearity---and so
is $\hat\sigma({1)}$ so $G_{jk}(t_1,t_0) = \delta_{jk}$ (full line). There is a
Fourier transform between $t_1$ and $t_2$ so $G_{jk}(t_2,t_1)=
\frac{1}{\sqrt\Omega} e^{i2\pi jk/\Omega}$ (double lines) according to
eq.(\ref{fourier}). Finally, by definition of the transformation
separating $t_2$ and $t_3$ [eq.(\ref{transfo})], the Green functions
represented by heavy lines are purely imaginary $\frac{i}{\sqrt 2}$ while those
represented by full lines are real $\frac{1}{\sqrt 2}$.

Griffiths has shown that when we are restricted to fine grained sets
of projectors, the medium consistency
condition eq.(\ref{medium}) is equivalent to the following. Given a vertex at time $0$ and a
vertex at time $t_f$, there is at most one path going forward in time
connecting them. Theorem \ref{fine} indicates that this formulation is
also valid for coarse grained sets. For example, we see that the
family of figure \ref{graphe_lemme} is not medium consistent since
there are two paths joining each end of the graph. These paths form
loops in the graph so the weak consistency condition is equivalent to
the absence of loops in the graph.

To serve the purpose of QIP, I have adapted this 
result to the weak consistency condition eq.(\ref{weak}).  

\begin{theorem}
The family of fine grained histories is weakly consistent
if the product of the Green function around all loops
in the associated graph is purely imaginary. [Note that
$G^{ij}(t_k,t_{k+1})  = (G^{ij}(t_{k+1},t_k))^*$. Also, $0$ is imaginary so
no loops yields consistency.]
\end{theorem}

\noindent
{\it Proof.} In order to prove this theorem, we only
need to point out that, when all histories are fine grained, the
product of the Green function around a loop is equal to the coherence
function for the two histories making up the loop. \hfill$\square$
\medskip

In the example of
figure \ref{graphe_lemme}, we can compute this product:
\begin{eqnarray} 
&&\frac{1}{\sqrt\Omega}\exp\left\{\frac{i2\pi jk}{\Omega}\right\} 
\left(\frac{1}{\sqrt 2}\right)\left(\frac{i}{\sqrt 2}\right)^* \nonumber\\
&\times& \frac{1}{\sqrt\Omega}
\left(\exp\left\{\frac{i2\pi j(k+\Omega
      /2)}{\Omega}\right\}\right)^* \\
&=& \frac{i}{2\Omega} \nonumber 
\end{eqnarray} 
which is purely imaginary and so weakly consistent.
It would be of great interest if this
technique could be generalized to the computing-consistency
condition. 

Theorem \ref{fine} applies only to systems with an initial pure
state. Hence, if the initial state of the computer is mixed, looking
among the fine grained measurements may not be applicable. Nevertheless, if
the initial state is pseudopure \cite{CFH1997} as one usually has
in NMR quantum computing \cite{CLKVHBBFLMNPSTWZ2000}, one can use the associated pure
state to verify the consistency.

\begin{lemma}
Let ${\mathcal S}=\{\ket{\psi}\bra{\psi},
   \sigma^{(1)}, \ldots , \sigma^{(n)}\}$ be a family of CH's
(any consistency condition will do).
Then ${\mathcal S'}=\{\rho, \sigma^{(1)}, \ldots , \sigma^{(n)}\}$
is computing-consistent with
\begin{equation}
\rho=\frac{1-\nu}{\Omega}\II +\nu \ket{\psi}\bra{\psi}.
\label{pseudo}
\end{equation}
\end{lemma}

\noindent {\it Proof}. 
To prove this, it is general to assume that the family ${\mathcal S}$ is
computing-consistent [eq.(\ref{hierarchy})]. It is then straightforward
to verify that ${\mathcal S'}$ satisfies eq.(\ref{comp}). \hfill$\square$
\medskip

As the last example illustrates, graphs can be very useful to analyze
the classicality of a
circuit but are unfortunately not always applicable. Even if the graphs can
be used to analyze the weak consistency, there might exist weak
consistent families that cannot be fine grained, and hence that cannot be
constructed from a graph. It was also postulated by Gell-Mann and
Hartle that there exist medium consistent families that cannot be fine
grained\footnote{Here, the word decoherent is used as a synonym of
  consistent: ``{\it
  Completely fine-grained histories cannot be assigned probabilities;
  only suitable coarse-grained histories can.}'' \cite{GH1990},
``{\it Except for pathological cases, coarse-graining is necessary
  for decoherence.}'' \cite{GH1995}}; from our analysis, such systems must initially be in mixed
states. I believe that systems with initial mixed states can always be
described with fine grained histories if feedback is allowed but I
have not found a proof. The intuition behind this comes from the fact
that a mixed state can be purified with ancilla states and that
branch-dependent CH's can be described without feedback if one has
access to a 
larger Hilbert space.

\section{Discussion}
\label{results}

In this section, two types of conclusion will be drawn from the
application of the CH
formalism to QIP. The first type concerns the consequences that a complete
classicality analysis would have on quantum computing; it mainly
consists of novel error prevention techniques. Since such a complete
analysis is at this time infeasible, these considerations are mostly
speculative. The second type proposes new techniques to address
fundamental questions concerning the computational power of quantum
mechanics. Since these questions are purely theoretical, the fact that
no systematic consistency analysis yet exists will not affect the
validity of our conclusions. Unfortunately, the lack of such an
analysis will be reflected by the simplicity of our examples which,
nonetheless, clearly illustrate the new concepts.

\begin{figure}[!b]
\centering
\includegraphics[scale=0.4]{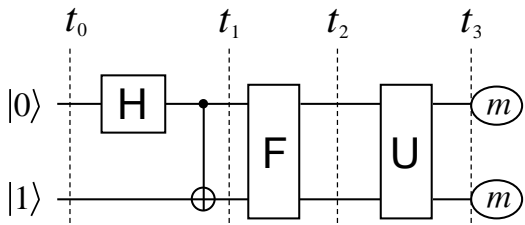}
\caption{Quantum circuit; the gates are described in the text.}
\label{circuit}
\end{figure}

We begin by showing that nontrivial consistent extensions exist. Consider the
quantum circuit of fig.\ref{circuit}
where $\mathsf{H}$ is the Hadamard gate, $\mathsf{F}$ is the $2$
qubits quantum Fourier transform, $\mathsf{U}$ is defined by
\begin{equation}
{\mathsf U} : \left\{
\begin{array}{lc}
\ket{00} \mapsto \ket{00} \\
\ket{01} \mapsto \left(\ket{01}+\sqrt{2}\ket{10}+\ket{11}\right)/2 \\
\ket{10} \mapsto \left(\ket{01}-\sqrt{2}\ket{10}+\ket{11}\right)/2 \\
\ket{11} \mapsto \left(\ket{01}-\ket{11}\right)/\sqrt{2}
\end{array}
\right.
\label{u}
\end{equation}
and the last measurements are made in the $0-1$ basis. A trivial
consistent extension of this family would consist of a
measurement that distinguishes the four Bell states 
\begin{equation}
\begin{array}{ll}
\ket{\Phi^{\pm}} = \frac{1}{\sqrt{2}}(\ket{00} \pm \ket{11}) \\
\ket{\Psi^{\pm}} = \frac{1}{\sqrt{2}}(\ket{01} \pm \ket{10})
\end{array}
\label{bell}
\end{equation}
at time $t_1$. This would obviously leave the final
result unchanged because at time $t_1$ the system is in the state
$\ket{\Psi^+}$. This measurement does not contribute to the
classicality of the circuit because it is not local so the quantum information
could not be substituted by classical information: entangled
states cannot be represented on classical spins. 

\begin{figure}[h!]
\begin{center}
\begin{picture}(210,140)
\put(30,20) {\circle*{5}}
\put(30,50) {\circle*{5}}
\put(30,80) {\circle*{5}}
\put(30,110) {\circle*{5}}
\put(80,20) {\circle*{5}}
\put(80,50) {\circle*{5}}
\put(80,80) {\circle*{5}}
\put(80,110) {\circle*{5}}
\put(130,20) {\circle*{5}}
\put(130,50) {\circle*{5}}
\put(130,80) {\circle*{5}}
\put(130,110) {\circle*{5}}
\put(180,20) {\circle*{5}}
\put(180,50) {\circle*{5}}
\put(180,80) {\circle*{5}}
\put(180,110) {\circle*{5}}
\put(30,80) {\line(1,0){50}}
\put(80,80) {\line(5,3){50}}
\put(130,110) {\line(1,0){50}}
\put(130,80) {\line(1,0){50}}
\put(130,80) {\line(5,-3){50}}
\put(130,80) {\line(5,-6){50}}
\put(130,20) {\line(5,6){50}}
\put(130,20) {\line(1,0){50}}
\put(80,79.4) {\line(1,0){50}}
\put(80,80.6) {\line(1,0){50}}
\put(80,79) {\line(5,-6){50}}
\put(80,81) {\line(5,-6){50}}
\put(15,20) {\makebox(0,0)[r]{$3$}}
\put(15,50) {\makebox(0,0)[r]{$2$}}
\put(15,80) {\makebox(0,0)[r]{$1$}}
\put(15,110) {\makebox(0,0)[r]{$\alpha=0$}}
\put(30,120) {\makebox(0,0)[c]{$\rho$}}
\put(80,120) {\makebox(0,0)[c]{\large$\hat\sigma^{(1)}$}}
\put(130,120) {\makebox(0,0)[c]{\large$\hat\sigma^{(2)}$}}
\put(180,120) {\makebox(0,0)[c]{\large$\hat\sigma^{(3)}$}}
\end{picture}
\caption{Graph analysis of the quantum circuit of fig.\ref{circuit}. The
  products of the Green functions around the loops are imaginary;
  hence, ${\mathcal
    S}=\{\ket{0}\bra{0},\sigma^{(1)},\sigma^{(2)},\sigma^{(3)}\}$ is
  consistent.} 
\label{graph}
\end{center}
\end{figure}
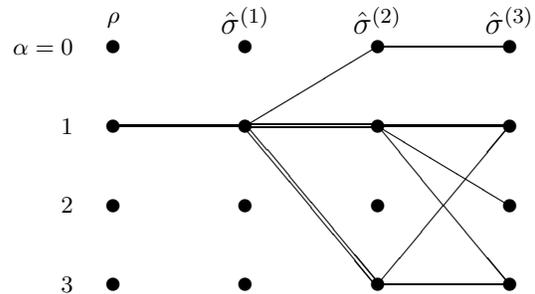

A nontrivial local consistent extension is given by measuring
both qubits in the computational basis at time $t_2$ (this defines $\sigma^{(2)}$). To convince
oneself, one should consider the graph of fig.\ref{graph} constructed from the
family ${\mathcal S}=\{ \ket{01}\bra{01},
   \sigma^{(1)},\sigma^{(2)},\sigma^{(3)}\}$ where we have kept the Bell measurement
$\sigma^{(1)}$ for sake of analysis (if ${\mathcal S}$ is consistent with
$\sigma^{(1)}$, it is also without since this is coarse graining). 
Also for sake of clarity, we have indicated only the Green functions that
are connected in some way to $\ket{\phi^{(0)}_2}=\ket{01}$ since they
are the only ones that might cause problems. There are two loops in
this graph, but the product of
the Green functions around these loops is imaginary. Indeed, all the
Green functions are real except $G_{11}(t_2,t_1) = \frac{-1+i}{\sqrt 8}$ and 
$G_{13}(t_2,t_1) = \frac{-1-i}{\sqrt 8}$ which are indicated by double
lines on the graph; furthermore, $G_{11}(t_2,t_1)G_{13}(t_2,t_1)^* = \frac{-i}{4}$; hence the 
measurement $\sigma^{(2)}$ is a weakly consistent extension.

To understand what is going on, it is convenient to
write the state of the system at time $t_2$: $\ket{\psi(t_2)} =
   [2\ket{00} + (-1+i)\ket{01} +(-1-i)\ket{11}]/\sqrt{8}$. This state has
 0.4165 $ebits$ of entanglement that are destroyed by the
measurement $\sigma^{(2)}$. Hence, this demonstrates that it is
possible to destroy 
entanglement in a quantum computer without affecting the final
result. One could argue that the final measurement will destroy
entanglement so $\sigma^{(2)}$ is only doing what would later be done by
$\sigma^{(3)}$. This is the case of the semiclassical quantum Fourier
transform of Griffiths and Niu \cite{GN1996},
where the final measurements were performed ahead of time, thus allowing the
substitution of QIP by CIP. This is not the case here. To convince
oneself, apply $\mathsf U^{-1}$ to the states of the final
measurement---the computational states---and verify that the result is
entangled. Therefore, in both directions of time, we are destroying
entanglement without affecting the final result. 

The skeptics can be convinced that the measurement is indeed
consistent by verifying that the probability sum rules are
satisfied. [Recall that the weak consistency condition that
was verified with the help of the graph of fig.\ref{graph} is stronger
than the computing consistency condition eq.(\ref{hierarchy}) which
itself implies the validity of the classical probability sum rules
eq.(\ref{stoch}).] To do this, construct the transition probability matrix
${\mathcal T} = (t_{ij})$ where $t_{ij}$ is the probability that the
system is measured in state $i$ at time $t_3$ after it is observed in state
$j$ at time $t_2$ as in eq.(\ref{stoch}). Obviously, $t_{ij} =
|u_{ij}|^2$, the square norm 
of elements of matrix $\mathsf U$ of eq.(\ref{u}).
If $P_i(t_k)$ denotes the probability of observing the system in state
$i$ at time $t_k$ [{\it i.e.}, $P_i(t_k) = |\bracket{i}{\psi(t_k)}|^2$ ],
 the probability sum rules should read $P_i(t_4) =
\sum_j t_{ij} P_j(t_3)$. Indeed, we get
\begin{equation}
\left(
\begin{array}{c}
1/2 \\
3/16 \\ 
1/8 \\ 
3/16
\end{array}
\right) =\left( 
\begin{array}{cccc}
1 & 0 & 0 & 0 \\ 
0 & 1/4 & 1/4 & 1/2 \\ 
0 & 1/2 & 1/2 & 0 \\ 
0 & 1/4 & 1/4 & 1/2
\end{array}
\right) \left( 
\begin{array}{c}
1/2 \\ 
1/4 \\ 
0 \\ 
1/4
\end{array}
\right) 
\end{equation}
as claimed. Therefore, the effect of gate $U$ {\it as seen from
  the final measurement} can be simulated by a stochastic classical
model. 

\subsection{Classically controlled operations}

At first sight, forcing the state of the computer to be classical at
time $t_2$ might not seem so exciting. Nevertheless, Dowker and Kent
have shown that if a measurement $\sigma$ appears twice in a
consistent family, say at times $t_a$ and $t_b$ with $t_a<t_b$, then it can
be repeated anywhere between $t_a$ and $t_b$ \cite{DK1996}. Let me
translate this 
result in terms of quantum computing: If qubit $k$ ($q_k$) can be made
classical at time $t_a$ and all operations between $t_a$ and $t_b$
are diagonal in the consistent basis of $q_k$,
then all quantum operations between $t_a$ and $t_b$ can be replaced by
classically 
controlled operations. This is the generalization of the observation
by Griffiths and Niu \cite{GN1996}
 that measuring a control qubit at the end of
the computation is equivalent to measuring it before the controlled operations.

\subsection{Noise robustness}

The classicality analysis can be used to enhance the noise robustness of a
computer for a given algorithm. Assume that $q_k$ can consistently
be measured at time $t_j$. Obviously, performing this measurement would
protect it against decoherence (uncontrolled measurement by an
exterior environment) since its information could be encoded
classically. Unfortunately, projective measurements cannot be
performed in a NMR setting \cite{EBW1994}. Nevertheless, if the decoherence is in a known
local basis, there is still hope. One can apply a rotation on $q_k$ so
that
the consistent basis agrees with the decoherence basis (rotating $q_k$
is like changing the basis in which decoherence occurs). Hence,
decoherence will perturb the state of the computer but in a
consistent way. This procedure can be seen as bringing the information
of the quantum
computer into a decoherence free-subspace;
 its purpose is to protect not
the state but the outcome of the computation.
For example, if known local decoherence takes place between gates
$\mathsf{F}$ and $\mathsf{U}$ of fig.\ref{circuit}, rotating both
qubits will protect the output of the computation, no matter what
state the computer is in at that time.

Now assume that the decoherence at time $t_2$ is still local but in an
unknown basis. Entanglement will then be destroyed in an uncontrolled
fashion and the refocusing scheme is of no help. Intuition tells us
that the damage should be less if we first destroy the entanglement in
a consistent fashion and then let decoherence perturb the system. In
other words, before local decoherence takes place, we force the system
into a local state, hoping to reduce its effect. (This is like
preventing forest fires by burning them down!) This intuition can be
verified using our toy model of fig.\ref{circuit}. Decoherence will
result in a probability distribution $\tilde{P}$ over the final
outcomes that is different from the original unperturbed distribution
$P$. The error due to decoherence can be measured by the relative
entropy $H(P||\tilde{P})$ (Kullback-Leibler distance \cite{CT1991}).
This entropy is calculated in two situations: $H_0$
measures the error caused by decoherence in the absence of the
consistent measurement at time $t_2$; $H_m$ is the error due to
decoherence when the system is forced into local state at time
$t_2$. Since the local basis in which decoherence occurs is unknown, we
must average the entropies over all local bases---integrate over two
Bloch sphere surfaces---we obtain a reduction of relative entropy of
about $25\%$ when consistent measurements are applied. (A
statistician's $\chi^2$ approach gives very similar results.)

\subsection{Mixed states}

Whether mixed state quantum computing can increase one's
computational power over classical computing is a question 
of great interest.
The main objection to such an increase is that, when the state is
highly mixed, it can always be written as a mixture of unentangled
states. Hence, it is possible that a mixed state algorithm can be
decomposed into a mixture of pure state algorithms that are in local
states throughout the computation \cite{PP2001}. This would suggest
that the mixed state algorithm can be simulated efficiently by a
classical probabilistic computer.  On the other hand, recent work by
Knill and Laflamme \cite{KL1998} gives indications that the answer is
positive, but does not provide a proof. 

When the initial state of the quantum computer is pure, the absence of
entanglement throughout the computation implies that all gates in the
circuit are classical and therefore can trivially be simulated on a
classical device \cite{BCJLPS1999}. When the initial state is highly
mixed, even fundamentally quantum operations will not create
entanglement. At first sight, this seems to imply that these
systems behave classically. Here, I show that, while the state of the
quantum computer is stroboscopically classical, no classical dynamics
can logically explain its evolution.

Consider a state of the the form eq.(\ref{pseudo}). For a two-event
family, the coherence function eq.(\ref{coherence}) reads
\begin{eqnarray}
&&Tr\left\{P^{(2)}_{\alpha_2}P^{(1)}_{\alpha_1}\rho
P^{(1)}_{\beta_1}P^{(2)}_{\beta_2}\right\} \nonumber\\
&=&\frac{(1-\nu)}{\Omega}
\delta_{\alpha_1\beta_1}\delta_{\alpha_2\beta_2}
Tr\left\{P^{(2)}_{\alpha_2}P^{(1)}_{\alpha_1}\right\} \\
&+&\nu\delta_{\alpha_2\beta_2}
Tr\left\{P^{(2)}_{\alpha_2}P^{(1)}_{\alpha_1}
\ket{\psi}\bra{\psi}P^{(1)}_{\beta_1}\right\}\nonumber
\end{eqnarray}
so the family is consistent if and only if the family with initial pure state
$\ket{\psi}$ is consistent. To complete the argument, choose your
favorite two-event family whose evolution cannot be explained in a
local fashion when the initial state is pure (I suggest Young's slit
experiment). This does not answer the question concerning the
computational power of mixed states but it indicates that there is
something fundamentally quantum in their evolution, so the main
objection does not hold. 

In fact, one can build a strong intuition about this result using
Young's slit experiment. Assume that the experiment is performed using
a ``pseudocoherent'' light source. For example, one could point a
laser and a regular (incoherent) light at the slits both at the same
time. Without the regular light, 
one observes the usual interference pattern. This implies that most
photons 
from the laser go through both slits. When the regular light is added,
the observed pattern is the classical superposition of a smooth
pattern (no interference) and the original interference
pattern. This cannot be explained without admitting that some photons
still go through both slits. Hence, even if a great number of photons
are not in a coherent phase, some fundamentally quantum phenomenon is
still taking place.

\subsection{Uniformity}

When analyzing a quantum circuit, one should be concerned with the
notion of uniformity. A quantum algorithm is represented by a family of
quantum circuits, one for each input size. The construction rule for
circuit of size $\ell+1$ given the one of size $\ell$ must be
simple. Therefore, a consistency analysis should be uniform, otherwise
it is useless in practice. The semiclassical quantum Fourier
transform of Griffiths and Niu is an example of a uniform consistency
analysis. 

From this point of view, the analysis of the circuit shown in figure
\ref{circuit} is not quite illuminating since it does not belong to a
family of circuits. Nevertheless, this example was not intended as a
practical one, but was used to illustrate a crucial point: the dynamics
generated by gate $\mathsf{U}$ on state $\ket{\psi(t_2)}$, while
highly nonclassical for the reasons mentioned earlier, is classical
{\it as 
  seen from the measurements}. In other words, there are fundamentally
quantum effects happening in the circuit, but the fixed final
measurement is blind to this quantumness. This would be impossible
if we required the extensions to be consistent for all choices of
final measurement; fortunately, the final measurements are fixed in a
quantum circuit.

The notion of uniformity allows us to weaken the locality restriction
on the sets of projectors. By restraining the measurement to single
qubit, we can simulate the {\it effective} dynamics of the quantum
system by classical spins, or equivalently by keeping track of two angles
and a radius of a Bloch sphere vector per qubit. By going to higher dimensional classical
simulators, we can perform joint measurements on many
qubits. Nevertheless, in order to have the size of the classical
simulator growing polynomially, one should restrict the size of these
joint measurements to be logarithmic. Here the notion of uniformity
is crucial. 

This is a generalization of the result of Aharonov and Ben-Or
\cite{AB1996} mentioned in the introduction. They established
that the absence of large scale coherence in a quantum system allows
an efficient classical simulation.
Here I show that it is not the absence of
coherence that is crucial but the absence of {\it effective}
coherence---the one that affects the coherence function.

\section{Conclusion}
\label{conclusion}

I have shown that not all the information processed by a
quantum computer is required to be quantum for the success of the
algorithm. The CH formalism together with an extra condition indicates
how and when one can substitute quantum information by classical
information. Classicality has helped increase the noise robustness of
a quantum computer. It has also given new
indications on the controversial question of computing with mixed
states. 

Quantum systems seem to be hard to simulate classically. Feynman's
argument---the exponentially growing size of the matrices---misses an important
point: only a linear fraction of this information is accessible by
fixed measurement. To be fair, classical simulation of quantum systems
should be concerned only with effective dynamics---the dynamics seen by
the measurements. The CH formalism is the appropriate tool to describe
this dynamics.

Classicality also allows the use of different
implementations of quantum computers within the same
computation. Since the information conveyed by the computer is made
classical at different stages of the computation, it can easily be
transferred to a different system. This new possibility might also have
implications in multiparty quantum computing.

The most exciting feature of classicality is that a complete CH analysis
of an algorithm would pinpoint what is fundamentally quantum in it,
hence answering the question ``What gives the extra computing
power to quantum mechanics, if any?'' Unfortunately, a complete analysis is
at this time impossible since the right tools have not been found. The
results of section 3 and \cite{DK1996} provide new hints in this
direction but further studies are required. It is encouraging to
notice that what is probably the most promising tool---feedback---has not
even been explored up to now.

\section*{Acknowledgments} 

\medskip\noindent
I am grateful to Gilles Brassard for inspiring conversations and useful
comments on this manuscript. 
I would also like to thank Alexandre Blais, Raymond Laflamme, Lorenza
Viola and Wojciech \.Zurek for stimulating and enjoyable discussions. 
This work was supported by Canada's NSERC.

\end{document}